\documentclass[onecolumn]{revtex4-2}

\usepackage{amsmath,amssymb,graphicx,hyperref,float}

\usepackage{xcolor}
\usepackage{hyperref}
\hypersetup{
colorlinks,
linkcolor={red!50!black},
citecolor={blue!80!black},
urlcolor={blue!80!black}
}

\newcommand{\ket}[1]{\ensuremath{\left| #1 \right\rangle}}
\newcommand{\bra}[1]{\ensuremath{\left\langle #1 \right|}}

\begin{document}

\title{Majorana excitons in a Kitaev chain of semiconductor quantum dots in a nanowire}

\author{Mahan Mohseni, Hassan Allami, Daniel Miravet, David J. Gayowsky, Marek Korkusinski*, Pawel Hawrylak}
\affiliation{Department of Physics, University of Ottawa, Ottawa, ON K1N 6N5, Canada}

\altaffiliation{Security and Disruptive Technologies, National Research Council, Ottawa, Canada K1A0R6}

\begin{abstract}
We present here a theory of Majorana excitons, photo-excited  conduction electron-valence band hole pairs, interacting with Majorana Fermions in a Kitaev chain of semiconductor quantum dots embedded in a  nanowire. Using analytical tools and exact diagonalisation methods we identify the presence of Majorana Zero Modes in the nanowire absorption spectra.
\end{abstract}

\maketitle

\section{Introduction}

There is currently interest in realizing synthetic topological quantum matter with topologically protected quasiparticles at its edges   \cite{gyongyosi2019survey,field2018introduction,campbell2017roads}, with potential application in topological quantum computation
\cite{stern2013topological,nayak2008non,sarma2015majorana,das2006topological,
freedman2003topological,jaworowski2019quantum}.
Haldane fractional spin quasiparticles in a spin one chain and Majorana Fermions in topological superconductors are good examples \cite{haldane1983nonlinear,kitaev2001unpaired,jaworowski2019quantum}.
To realize Majorana Fermions  Kitaev proposed \cite{kitaev2001unpaired,kitaev2003fault} a chain of quantum dots on a  p-wave superconductor that carries such non-local zero energy Majorana Fermions localized on its two ends, the Majorana zero modes (MZMs).
Since then there have been numerous proposals to realize the Kitaev chain \cite{lutchyn2010majorana,mourik2012signatures,s-type_sc_qd_array,poor_man_mzm,minimal_mzm,nadj2014observation,sun2017detection}.
In all cases, experimental confirmation of the presence of the MZMs has proved to be a non-trivial and challenging task \cite{liu2011detecting,jack2021detecting,pientka2013signatures,pikulin2021protocol,liu2012zero,sarma2021disorder,rubbert2016detecting,aghaee2022inas,baldelli2022detecting}.

Recent progress in semiconductor quantum dots in nanowires \cite{cygorek2020atomistic,Jmanalo,PhysRevApplied.14.034011,doi:10.1021/nl303327h,jaworowski2017macroscopic,phoenix2022magnetic,northeast2021optical,laferriere2021systematic} opens the possibility of realizing Kitaev chains and optical detection of their Majorana zero modes.
In this work, we  consider such an array of InAsP quantum dots embedded in an InP nanowire as the material system \cite{cygorek2020atomistic,Jmanalo,PhysRevApplied.14.034011,doi:10.1021/nl303327h,jaworowski2017macroscopic,phoenix2022magnetic,northeast2021optical,laferriere2021systematic} for realization of MZM, and study its signature in light-matter interaction.
As the schematic in Fig.~\ref{fig:schem_ham}(a) shows, we combine a semiconductor nanowire with p-wave superconductor \cite{talantsev2019p,wang2019theory,yuan2014possible,frigeri2004superconductivity,hardy2005p,ishida1998spin}. The p-wave pairing in this system is introduced by proximity effect among electrons that are spin-polarized by an external magnetic field, making sure that Cooper pairs can only form between electrons in the conduction band (CB) of adjacent dots.
We will show that one can tune the system parameters into topological regime, where two MZMs appear at the two ends of the chain.
 With semiconductor quantum dots, light can generate a hole in the valence band (VB) and an electron in the conduction band. The electron adds to an existing gas of Majorana Fermions while
the hole then  interacts with all the quasiparticles of the Kitaev chain, including MZMs, to form composite objects similar to excitons and trions in the Fermi Edge Singularity problem \cite{mahan2012many,PhysRevB.51.10880,HAWRYLAK1992525,PhysRevB.44.3821}. This leads to a structure in the absorption spectrum of the chain as a function of photon energy. Here we present a theory for the signatures of the MZMs in the optical spectra of the semiconductor nanowire.

After describing the model in Section~\ref{sec:description}, in Section~\ref{sec:kitaev} we introduce the exact diagonalization (ED) method, and introduce Majorana and bond Fermion   representation of the Kitaev Hamiltonian.
Next, in Section~\ref{sec:one_hole} we describe exciton-Majorana Fermion complexes and predict the absorption spectrum. We focus discussion on the optical signature of the MZM in the absorption spectrum.
Finally, in Section~\ref{sec:conclusion} we conclude by summarising our results and discuss potential experiments detecting Majorana Fermions in a semiconductor Kitaev chain.

\section{Kitaev chain in a semiconductor nanowire \label{sec:description}}

\begin{figure}[h]
    \centering
	\includegraphics[width=\columnwidth]{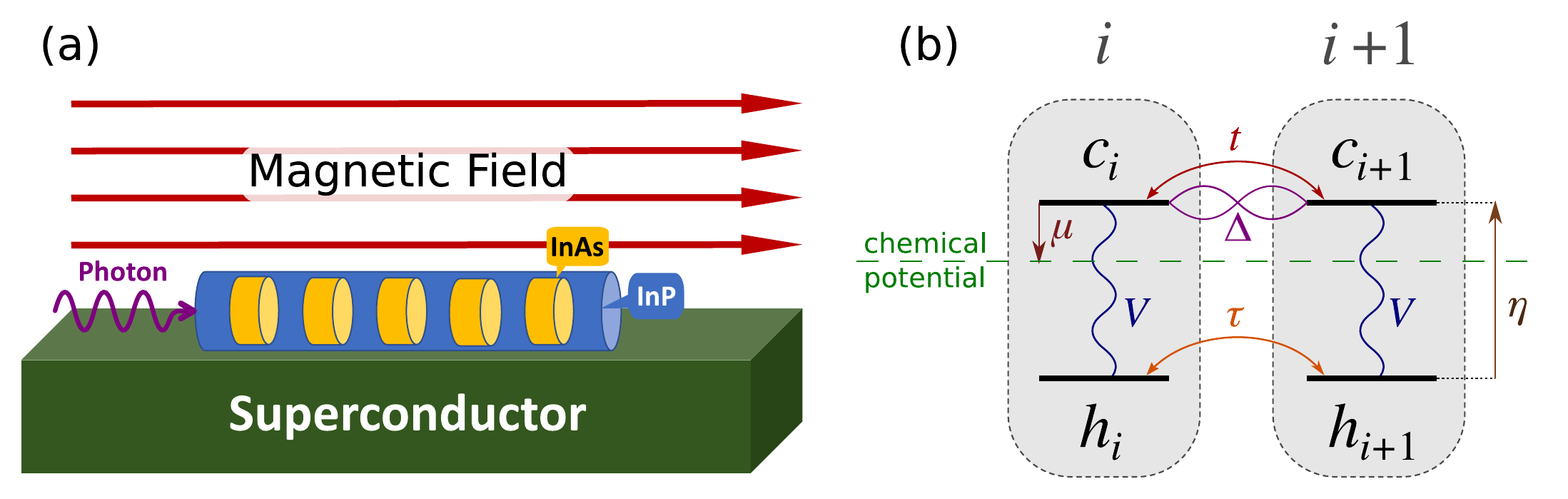}
	\caption{(a) Schematic of the system and the light absorption experiment. (b) Schematic of the Hamiltonian terms between two adjacent dots according to \eqref{eq:full_Ham}, where conduction(valence) levels are labeled by $c_i(h_i)$ operators. The conduction level is the reference of energy, hence the downward arrow indicates negative $\mu$.}
	\label{fig:schem_ham}
    \end{figure}
 Figure~\ref{fig:schem_ham}(a) shows a schematic representation of the  Kitaev chain  we are considering. It consists of a hexagonal InP nanowire with an array of embedded InAsP quantum dots in the proximity of a p-wave superconductor \cite{talantsev2019p,wang2019theory,yuan2014possible,frigeri2004superconductivity,hardy2005p,ishida1998spin}, in the presence of applied external magnetic field.
 Such arrays have been extensively investigated \cite{cygorek2020atomistic,Jmanalo,PhysRevApplied.14.034011,doi:10.1021/nl303327h,jaworowski2017macroscopic,phoenix2022magnetic,northeast2021optical,laferriere2021systematic}, including their excitonic complexes \cite{cygorek2020atomistic,laferriere2021systematic}.
 As Fig.~\ref{fig:schem_ham}(b) shows, in our model we include the lowest conduction spin level of each dot and the highest spin valence band level, which are effectively both spin-polarized due to the external magnetic field.
 Consequently, in the presence of superconductivity, only the electrons from the adjacent conduction levels can pair up, as there is only one conduction level available in each dot. 
 The Kitaev Hamiltonian $H_e$ in \eqref{eq:Kitaev_Ham} describes the hopping and pairing of electrons in conduction band levels.
 The chemical potential  is tuned to bring the chain to near half filling in the absence of superconductivity.
Therefore, in the equilibrium the valence levels are full and the system is described by the Kitaev Hamiltonian.
 However,  when a photon with energy close to the band gap of InAsP illuminates the dots, it generates a hole in VB and an electron in CB. The electron becomes one of the electrons in CB and decomposes into quasiparticles of the superconducting state. The hole then forms a bound state with the quasiparticles of the electronic system that is in a collective superconducting state. These possible bound states generate peaks in the absorption spectrum of the system, among which there is the signature of MZM, as we shall show below.
 
The hole is described by a simple tight binding Hamiltonian $H_h$ in \eqref{eq:hole_Ham}. We also consider electron-hole interaction, $H_{int}$ in \eqref{eq:eh_int_Ham}, which is strongest when both conduction electron and valence band hole are on the same quantum dot.
 Hence, we write the full Hamiltonian of the system as

    \begin{subequations}
    \label{eq:full_Ham}
    \begin{align}
    & H = H_e + H_h + H_{int},\nonumber\\
    	\label{eq:Kitaev_Ham}
     & H_e = t\sum_{i=1}^{N-1} c^\dagger_{i+1} c_i +h.c. + \Delta \sum_{i=1}^{N-1} c^\dagger_{i+1}c^\dagger_i +h.c.
		-\mu \sum_{i=1}^{N} c^\dagger_{i}c_i,\\
  \label{eq:hole_Ham}
  & H_h = - \tau \sum_{i=1}^{N-1} h^\dagger_{i+1} h_i + h.c. + \eta \sum_{i=1}^{N} h_i^\dagger h_i, \\
  \label{eq:eh_int_Ham}
  & H_{int} = -V \sum_{i=1}^N n_i^e n_i^h,
    \end{align}
    \label{eq:bands}
    \end{subequations}
where $c_i^\dagger(h_i^\dagger)$ is the normal Fermionic creation operator of an electron(hole) in dot $i$, $t(\tau)$ is hopping between adjacent conduction(valence) levels, $\Delta$ is pairing energy between adjacent conduction levels, $\mu$ is the chemical potential measured from the conduction energy level, and $\eta$ is the CB to VB energy gap in each dot.
In the interaction term \eqref{eq:eh_int_Ham}, $V$ is the Coulomb attraction energy between electrons and holes, where we also introduced $n_i^e = c_i^\dagger c_i (n_i^h = h_i^\dagger h_i)$, the electron(hole) number operator in dot $i$.
Figure.~\ref{fig:schem_ham}(b) schematically shows different terms of \eqref{eq:full_Ham} between two adjacent dots.

    Next, before describing the  absorption experiment, we start with a brief discussion of Kitaev Hamiltonian.

\section{Majorana and bond Fermions in Kitaev Hamiltonian \label{sec:kitaev}}

The Kitaev Hamiltonian $H_e$ in \eqref{eq:Kitaev_Ham}, originally introduced in Ref.~\cite{kitaev2001unpaired}, supports two MZMs localized on the two ends of the chain, when the Hamiltonian is in topological regime.
For a finite chain, the  topological region is centered on parameters $\Delta=t$ and $\mu=0$, which is our focus throughout this work.
Here, after describing the exact diagonalization (ED) method for normal Fermions, following Kitaev \cite{kitaev2001unpaired}, we show how using Majorana Fermions reveals the usefulness of a new set of  Fermions  we refer to as \textit{bond Fermions}.
Next, after matching energy spectra obtained by ED in both normal and bond Fermion bases, we shall use the bond Fermion basis for the rest of the paper.

\subsection{Exact diagonalization in normal Fermion basis}

We start off by introducing the exact diagonalization method (ED) for finding the energy spectrum of the Kitaev Hamiltonian. In ED we span the Hilbert space of the system by configuration basis \cite{ED_chap_2008}. For our electronic system being made of $N$ spinless orbitals, there are ${N \choose 0} + {N \choose 1} + ... + {N \choose N - 1} + {N \choose N}=2^N$ possible configurations, which we construct as
    \begin{equation}
        \ket{\alpha_1 \dots \alpha_N} = \prod_{i=1}^N (c_i^\dagger)^{\alpha_i} \ket{0} ,
        \label{eq:kitaev_confs}
    \end{equation}
where $\ket{0}$ is the vacuum of electrons, $\alpha_i = 1 \text{ or } 0$, corresponds to having(1) or not having(0) electron in orbital $i$.
    
For a given number of electrons $M$ we generate electron configurations $p_M$.
But as Kitaev Hamiltonian, being a Hamiltonian for a superconductor, does not conserve particle number, its eigenstates are coherent linear combinations of electronic configurations with different electron numbers as
    \begin{equation}
        \ket{\psi^\nu} = \sum_{M,p_M} C^\nu_{M,p_M}\ket{{M,p_M}},
      \label{eq:ED_general_psi}
    \end{equation}
where we are populating $N$ sites with $M=0,1,\dots , N$ electrons. To solve for coefficients $C^\nu_{M,p_M}$, we apply the Hamiltonian on this state, and by using the orthogonality of the configurations we obtain the eigenvalue equation
\begin{equation}
      \sum_{p_M,M} \bra{q_{M'},M'} H_e \ket{p_M,M} C^\nu_{M,p_M} = E^\nu C^\nu_{M',q_{M'}}.
     \label{eq:Wfn_MatrixH}
\end{equation}
However, since the Kitaev Hamiltonian $H_e$ in \eqref{eq:Kitaev_Ham} only changes particle number in pairs,
the matrix element $\bra{q_{M'},M'} H_e \ket{p_M,M}$ is non-zero only if $M$ and $M'$ have the same parity, i.e. if they are both even or odd. This parity symmetry allows us to break the Hilbert space into two decoupled subspaces of even and odd configurations.
In Appendix~\ref{sec:N3} we explicitly show the configurations and the Hamiltonian matrix $\bra{q_{M'},M'} H_e \ket{p_M,M}$ in each of these subspaces, for the case of $N=3$.

\subsection{Bond Fermions}
\begin{figure}[h]
    \centering
	\includegraphics[width=0.8\columnwidth]{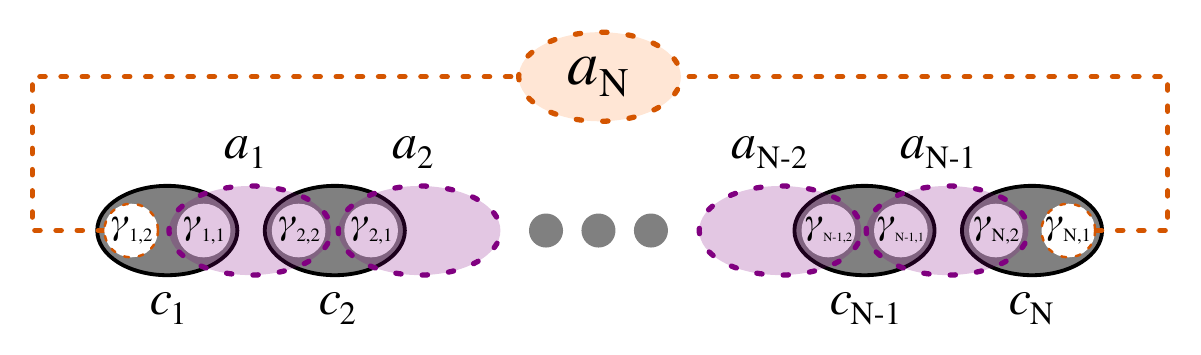}
	\caption{Schematic of Kitaev chain in the Majorana and bond representation, with non-zero bond Fermions in purple, and the nonlocal zero mode $a_N$ living on the two ends of the chain.}
	\label{fig:bond_Fermions}
    \end{figure}

We now express the Kitaev Hamiltonian in \eqref{eq:Kitaev_Ham} in terms of Majorana and bond Fermions.
First, as schematically shown in Fig.~\ref{fig:bond_Fermions}, we write each electron operator , $c$ and $ c^+ $ in terms of two Majorana Fermion operators $\gamma_{1}$ and $\gamma_{2}$ as
    \begin{align}
        \label{eq:c_to_gamma}
        & c_j = \frac{1}{2}(\gamma_{j,1} + i \gamma_{j,2}),
        & c_j^\dagger = \frac{1}{2}(\gamma_{j,1} - i \gamma_{j,2}),        
    \end{align}
    \label{eq:c_gamma_relations}
where the $\gamma$'s are Majorana Fermion operators.
Majorana Fermions
satisfy a slightly different anti-commutation relation than  the ordinary Fermions, 
$\lbrace\gamma_{i,\alpha},\gamma_{j,\beta}\rbrace=2\delta_{ij}\delta_{\alpha\beta}$.

Using \eqref{eq:c_to_gamma} and Majorana anti-commutation relations, the Hamiltonian $H_e$ can be written in terms of Majorana Fermions as
    \begin{equation}
     H_{e} = \frac i2 \left( (t+\Delta) \sum_{j=1}^{N-1}\gamma_{j,1}\gamma_{j+1,2} +  
     (t-\Delta) \sum_{j=1}^{N-1}\gamma_{j+1,1}\gamma_{j,2}
     - \mu \sum_{j=1}^{N} (\gamma_{j,1}\gamma_{j,2} - i)\right).
     \label{eq:Majorana_General_H}
    \end{equation}

The form in \eqref{eq:Majorana_General_H} shows the pairing between Majoranas of different types in adjacent sites. But it also shows that the Hamiltonian is not diagonal in Majorana Fermions, they are not quasiparticles of the Kitaev Hamiltonian.
Following Kitaev \cite{kitaev2001unpaired}, as shown in Fig.~\ref{fig:bond_Fermions}, we define a new set of Fermionic operators, bond Fermions, which are made of two Majoranas of different types from adjacent sites as
\begin{subequations}
\begin{align}
    \label{eq:aj_to_c}
    &a_j = \frac{1}{2}(\gamma_{j,1} + i\gamma_{j+1,2})= \frac{1}{2}(c_j^\dagger + c_j + c_{j+1} - c_{j+1}^\dagger),\\
    &a_N = \frac{1}{2}(\gamma_{N,1} + i \gamma_{1,2})
        = \frac{1}{2}(c_N^\dagger + c_N + c_{1} - c_{1}^\dagger),
    \label{eq:aN_to_c}
\end{align}
\label{eq:a_to_c}
\end{subequations}
where we also defined $a_N$,  to which we refer as the zero mode, out of the two unpaired Majoranas at the two ends of the chain, as shown in Fig.~\ref{fig:bond_Fermions}.
Then, the Hamiltonian in terms of bond Fermion operators is
\begin{align}
     H_{e} = \frac12 &\left(
     (t+\Delta) \sum_{j=1}^{N-1}(2a_j^{\dagger}a_j - 1) 
      + (t-\Delta) \sum_{j=1}^{N-1}
     (a_{j+1}^\dagger a_{j-1} + a_{j+1}a_{j-1} + h.c.)
     \right. \nonumber\\
        &\left. - \mu \sum_{j=1}^{N} 
         (1 + (a_{j}^\dagger a_{j-1} +  a_{j}a_{j-1} + h.c.) ) \right),
     \label{eq:General_H_BondFermions}
\end{align}
where in the second and the third sum one should identify $a_0\equiv a_N$.
Note that in topological regime, when $t=\Delta$ and $\mu=0$, the bond Fermions diagonalize the Hamiltonian in \eqref{eq:General_H_BondFermions} and reduce it to
\begin{equation}
     H_{e} = t \sum_{j=1}^{N-1}(2a_j^{\dagger}a_j - 1),
     \label{eq:bond_special}
\end{equation}
which implies a set of $N-1$ quasiparticles with energy $2t$, and one non-local quasiparticle $a_N$ with zero energy, hence the name zero mode.
In this case, since bond Fermions are the quasiparticles of Kitaev Hamiltonian, their configurations are the eigenstates of the system.

In this spirit, we also use bond Fermion configurations for exact diagonalization of Kitaev Hamiltonian.
In the same fashion as in \eqref{eq:kitaev_confs} we define bond Fermion configurations as
\begin{equation}
        \ket{\overline{\alpha_1 \dots \alpha_N}} = \prod_{i=1}^N (a_i^\dagger)^{\alpha_i} \ket{0_a},
        \label{eq:bond_confs}
\end{equation}
where $\ket{0_a}$ is the vacuum of bond Fermions, and  we used the \textit{overline} to distinguish these configurations from the normal Fermion configurations.
Next, an equation similar to the equation in \eqref{eq:ED_general_psi} can be written for the eigenstates of the Hamiltonian in terms of bond Fermion configurations, where now $\ket{M,p_M}$ would represent the $p_M$ configuration of having $M$ bond Fermions.
And similar to the case of normal Fermions, since $H_e$ in \eqref{eq:General_H_BondFermions} conserves the parity of bond Fermion numbers too, we can split the Hilbert space into even and odd subspaces.
In Appendix~\ref{sec:N3} we explicitly show the bond Fermion configurations and the Hamiltonian matrix of \eqref{eq:General_H_BondFermions} in each of these subspaces, for the case of $N=3$.
\subsection{Energy spectrum \label{sec:elec_energy_spec}}
To demonstrate the usefulness of bond Fermion basis, we now describe the energy spectrum of a chain of $N=3$ quantum dots, obtained both in the normal and bond Fermion basis.
    \begin{figure}[H]
    \centering
	\includegraphics[width=0.8\columnwidth]{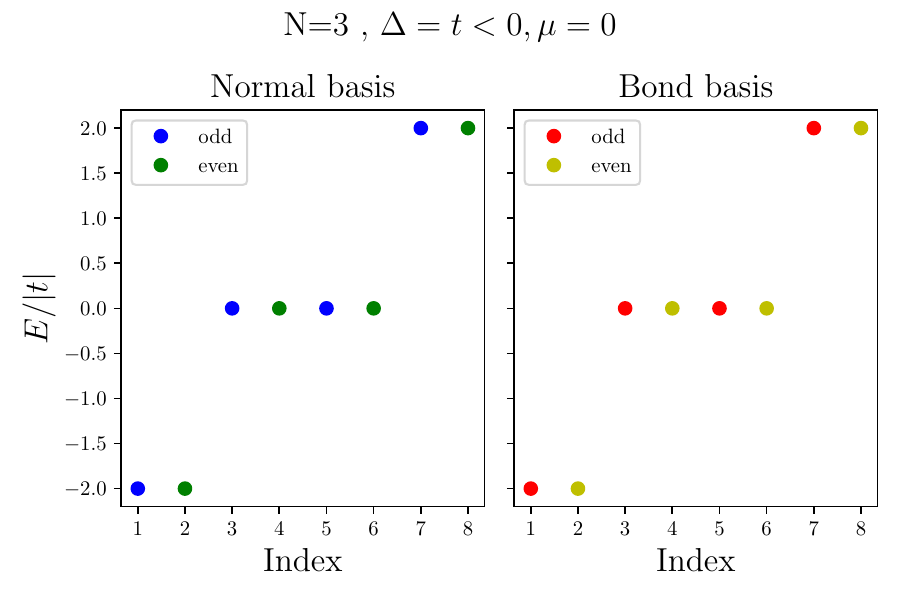}
	\caption{Energy spectra of Kitaev chain in normal(left) and bond(right) basis, where $\Delta = t<0$ and $\mu=0$. Energy is normalized to $|t|$.}
	\label{fig:energy_special}
    \end{figure}
Figure~\ref{fig:energy_special} shows the energy spectrum for the case of $\Delta=t<0$. Throughout the work, we consider $t<0$, as it is the case for conduction bands hopping integrals.
As we mentioned above, in this case, the configurations of bond Fermions are also the eigenstates of the system.
Being in topological regime, with these parameters the system has a doubly degenerate ground state, one in the odd subspace $\ket{\rm GS}=\ket{\overline{111}}$ with all bond Fermions, and the other in the even subspace $\ket{\overline{\rm GS}}=\ket{\overline{110}}$, which is missing the zero energy bond Fermion $a_N\equiv a_3$.
Next, we have the singly excited states, missing one non-zero bond Fermion, with excitation energy $2|t|$, from which we have two in each subspace, $\ket{a_1}=\ket{\overline{011}}$ and $\ket{a_2}=\ket{\overline{101}}$ in the even subspace, and $\ket{\overline{a_1}}=\ket{\overline{010}}$ and $\ket{\overline{a_2}}=\ket{\overline{100}}$ in the odd subspace.
Finally, in each subspace, there is one doubly excited state, missing two non-zero bond Fermion with excitation energy $4|t|$, $\ket{\overline{a_1a_2}}=\ket{000}$ in the even subspace, and $\ket{a_1a_2}=\ket{001}$ in the odd subspace.
Table~\ref{tab:elec_energy_spectrum} summarizes  the description of the spectrum in terms of bond Fermions.

\begin{table}[H]
	\caption{Describing the spectra plotted in Fig.~\ref{fig:energy_special}. Configurations of bond Fermions are the eigenstates of Kitaev Hamiltonian when $\Delta=t$ and $\mu=0$.\\}
	\label{tab:elec_energy_spectrum}
	\centering
	\begin{tabular}{c|cccccccc}
		index & 1 & 2 & 3 & 4 & 5 & 6 & 7 & 8  \\[5pt]
        \hline
		configuration & $\ket{\overline{111}}$ & $\ket{\overline{110}}$ & $\ket{\overline{010}}$ & $\ket{\overline{011}}$ & $\ket{\overline{100}}$ & $\ket{\overline{101}}$ & $\ket{\overline{001}}$ & $\ket{\overline{000}}$\\[10pt]
        label & $\ket{\rm GS}$ & $\ket{\overline{\rm GS}}$ & $\ket{\overline{a}_1}$ & $\ket{a_1}$ & $\ket{\overline{a}_2}$ & $\ket{a_2}$ & $\ket{a_1a_2}$ & $\ket{\overline{a_1a_2}}$\\[10pt]
		parity & odd & even & odd & even & odd & even & odd & even\\[7pt]
        \begin{tabular}{c}
             excitation \\ energy
        \end{tabular} & 0 & 0 & $2|t|$ & $2|t|$ & $2|t|$ & $2|t|$ & $4|t|$ & $4|t|$
	\end{tabular}
\end{table}

\section{Kitaev chain and a light induced valence hole \label{sec:one_hole}}
Absorption of a photon injects an electron-hole pair into the system. Therefore, the relevant optically excited states live in the subspace of all configurations with one hole. Here, after studying the energy spectrum of the full Hamiltonian in \eqref{eq:full_Ham} with one hole in the configuration space of bond Fermions, we discuss the absorption spectrum of the chain and the optical signature of the MZM. 
\subsection{Exact diagonalization of electron-hole system}
Having demonstrated the benefit of bond Fermion basis, we now study the Hamiltonian with one hole in the configuration basis of bond Fermions and one hole. Using $\ket{M,p_M; m}$ to refer to $M$ bond Fermions being in their $p_M$ configuration, and the hole being at site $m$, we can find the spectrum by solving an equation similar to \eqref{eq:Wfn_MatrixH}, but considering the full Hamiltonian $H$ in \eqref{eq:full_Ham} rather than $H_e$, and in the configuration basis of bond Fermions and one hole.

For instance, for $N=3$ dots, following the convention we introduced in Section~\ref{sec:elec_energy_spec} and Table~\ref{tab:elec_energy_spectrum}, we can list these configurations as
\begin{table}[H]
	\caption{Configurations of bond Fermions with one hole for $N=3$ dots.\\}
	\label{tab:bond_hole_configs}
	\centering
	\begin{tabular}{c|c}
        Even & Odd \\
        \hline\hline
        \begin{tabular}{cccc}
             $\ket{\overline{\rm GS};1}$ & $\ket{a_1;1}$ & $\ket{a_2;1}$ & $\ket{\overline{a_1a_2};1}$ \\[5pt]
             $\ket{\overline{\rm GS};2}$ & $\ket{a_1;2}$ & $\ket{a_2;2}$ & $\ket{\overline{a_1a_2};2}$\\[5pt]
             $\ket{\overline{\rm GS};3}$ & $\ket{a_1;3}$ & $\ket{a_2;3}$ & $\ket{\overline{a_1a_2};3}$
        \end{tabular}
        &\begin{tabular}{cccc}
            $\ket{{\rm GS};1}$ & $\ket{\overline{a}_1;1}$ & $\ket{\overline{a}_2;1}$ & $\ket{a_1a_2;1}$ \\[5pt]
             $\ket{{\rm GS};2}$ & $\ket{\overline{a}_1;2}$ & $\ket{\overline{a}_2;2}$ & $\ket{a_1a_2;2}$ \\[5pt]
             $\ket{{\rm GS};3}$ & $\ket{\overline{a}_1;3}$ & $\ket{\overline{a}_2;3}$ & $\ket{a_1a_2;3}$
        \end{tabular}
	\end{tabular}
\end{table}

In this subspace, the hole Hamiltonian $H_h$ in \eqref{eq:hole_Ham} amounts to a constant $\eta$ and mixes states with the same electronic configurations and different locations of the hole by hopping matrix element $\tau$.
Therefore, with the ordering in Table~\ref{tab:bond_hole_configs}, the full Hamiltonian with one hole for the example of $N=3$ dots has a structure like
\begin{equation}
H =
    \begin{bmatrix}
        H_1 & -\tau & 0 \\
        -\tau & H_2 & -\tau \\
        0 & -\tau & H_3 
    \end{bmatrix} + \eta,
    \label{eq:H_matrix_hole_structure_N3}
\end{equation}
where each block is a $4\times 4$ matrix, $\tau$ is the identity matrix times $\tau$, and the diagonal blocks are given by the matrix elements of $H_e + H_{int}$ in \eqref{eq:full_Ham} over the configurations in Table~\ref{tab:bond_hole_configs}.
The interaction term $H_{int}$ in \eqref{eq:eh_int_Ham} for each of the diagonal blocks $H_j$ is $-Vn_j^e$, and it mixes up different bond Fermion configurations as we have
\begin{subequations}
        \begin{align}
            \label{eq:ne_j_to_a_j}
            n_j^e &= \frac12 + \frac{1}{2}(a_{j-1}^\dagger a_j + a_{j-1}^\dagger a_j^\dagger +h.c.),
            & 1 < j \leq N ,\\
            \label{eq:ne1_to_a}
            n_1^e &= \frac12 + \frac{1}{2}(a_{N}^\dagger a_1 + a_N^\dagger a_1^\dagger +h.c.),
            & j=1,
        \end{align}
        \label{eq:ne_to_a}
\end{subequations}
which implies that when the hole is not at the two ends of the chain then the interaction mixes up two non-zero bond Fermions, and when it is at one of the two ends, the interaction mixes the zero mode with one of the non-zero ones. For instance, for the operator $n_1^e$ and $n_2^e$ in the even configuration basis in Table~\ref{tab:bond_hole_configs} we have
    \begin{align}
        &n_1^e =\frac12
        \begin{bmatrix}
            1 & -1 & 0 & 0 \\
            -1 & 1 & 0 & 0 \\
            0 & 0 & 1 & -1 \\
            0 & 0 & -1 & 1 \\
        \end{bmatrix},
        &n_2^e =\frac12
        \begin{bmatrix}
            1 & 0 & 0 & 1 \\
            0 & 1 & -1 & 0 \\
            0 & -1 & 1 & 0 \\
            1 & 0 & 0 & 1 \\
        \end{bmatrix},
        \label{eq:ne1_ne2_in_even_bond_config}
    \end{align}
while $n_3^e$, in a similar fashion to $n_1^e$, mixes $\ket{\overline{\rm GS}}$ with $\ket{a_2}$, and $\ket{a_1}$ with $\ket{\overline{a_1a_2}}$.

\subsection{Energy spectrum of the electron-hole system}

\begin{figure}[h]
    \centering
	\includegraphics[width=\columnwidth]{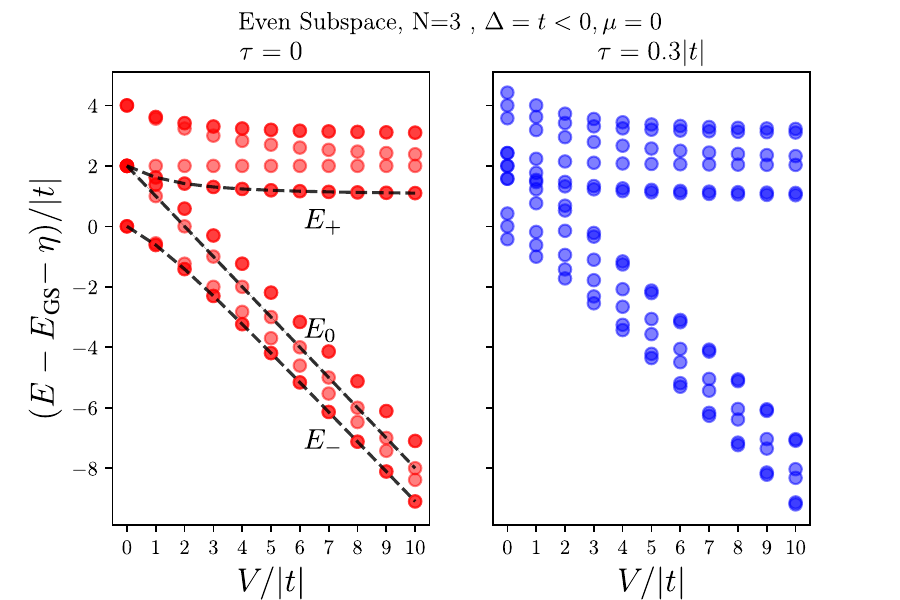}
	\caption{Energy spectra of the full Hamiltonian \eqref{eq:full_Ham} with one hole in the even subspace, as a function of electron-hole interaction $V$, for $N=3$ dots, $\Delta=t$ and $\mu=0$. (left) for the case of localized hole, $\tau=0$, (right) for a mobile hole with $\tau=0.3|t|$. The overlap of transparent markers makes the degenerate levels look darker. The peak energies $E_0$ and $E_\pm$  discussed in Section~\ref{sec:absorption_local_hole} are also shown according to Eqs.~(\ref{eq:E0} and \ref{eq:Epm}).}
	\label{fig:energy_1hole}
\end{figure}
Figure~\ref{fig:energy_1hole} shows the energy spectrum of a chain of length $N=3$ dots in the even subspace and for $\Delta=t$, and $\mu=0$, as the electron-hole interaction $V$ increases; for a localized hole ($\tau=0$) on the left panel, and for a mobile hole with $\tau=0.3|t|$ on the right panel. Both cases show branching into two groups, pertaining to bonding and antibonding pairs of states, mixed by the interaction $V$.

The case of localized hole allows us to understand the spectrum better. There are four states associated with each dot, and since the two end dots are geometrically the same, the spectrum always shows four pairs of doubly degenerate states.
As we show in the Appendix~\ref{sec:appendix},  and it can be seen from \eqref{eq:ne1_ne2_in_even_bond_config}, for the two end dots two of these four states are mixtures of $\ket{\overline{\rm GS};1(3)}$ and $\ket{a_{1(2)};1(3)}$ that give us visible peaks at $E_\pm$  described in Section~\ref{sec:absorption_local_hole}, and also indicated on the plot. The two other pairs of degenerate levels are mixtures of  $\ket{a_{2(1)};1(3)}$ and $\ket{\overline{a_1a_2};1(3)}$, which do not get excited by absorbing a photon.
For the middle dot, as can be seen from \eqref{eq:ne1_ne2_in_even_bond_config}, one pair of states are mixture of $\ket{a_1;2}$ and $\ket{a_2;2}$, where only the bonded state gets excited by absorbing a photon (see Appendix~\ref{sec:appendix}), resulting in the peak $E_0$; also described in Section~\ref{sec:absorption_local_hole} and shown on the plot. Finally, the last pair of states, which also do not get excited by absorbing a photon, are mixtures of $\ket{\overline{\rm GS};2}$ and $\ket{\overline{a_1a_2};1(3)}$.

As can be seen in the right panel of Fig.~\ref{fig:energy_1hole}, for a mobile hole, when $\tau\neq 0$, we still have four pairs of doubly degenerate states as a result of the chain's spatial symmetry. At $V=0$ for the case of localized hole there is an extra triple degeneracy because of the non-dispersive nature of the localized hole band. But for a mobile hole, it can be seen on the right panel of Fig.~\ref{fig:energy_1hole}, that at $V=0$ the degenerate levels split into sets of triples, corresponding to the three propagating modes of the hole band. More importantly, in this case, since the three dot subspaces are connected by hole hopping (see \eqref{eq:H_matrix_hole_structure_N3}), the above described pairs of states mix up by $\tau$, and the ones that are closer in energy mix more. As a result of this mixture, more peaks arise in the absorption spectrum, as we discuss in the next section.

\subsection{Absorption spectrum}
As the schematic in Fig.~\ref{fig:schem_ham}(a) shows, in absorption experiment a photon probes the chain along the nanowire.
InAsP dots having significantly smaller bandgap than InP bulk of the nanowire \cite{adachi_ch6}, guaranties that the photon can only be absorbed by the dots.
For calculating the absorption spectrum of the chain, we assume that the photon creates an electron-hole pair with uniform probability along the nanowire, and so define the polarization operator as

\begin{equation}
    P = \frac{1}{\sqrt{N}}\sum_{i=1}^N c_i^\dagger h_i^\dagger = \frac{1}{\sqrt{N}}\sum_{i=1}^N P_i,
    \label{eq:polarization_op}
\end{equation}
where we also introduced the local electron-hole pair operator $P_i=c_i^\dagger h_i^\dagger$.

We are assuming that one can also setup the system to create the electron-hole pair on a chosen specific  dot $i$ \cite{northeast2021optical,Dalacu2021Tailoring,laferriere2020multiplexed,laferriere2023position}, i.e. acting with the operator $P_i$ on the chain, rather than $P$.
As we discuss, having access to such spatially resolved spectrum is important in detecting the optical signature of the MZM.

The polarization operator $P_{(i)}$ -- we use this notation to simultaneously refer to $P$ and $P_i$ -- takes the ground state of the system to an excited state with one hole and one electron in CB. Since the ground state can be degenerate,  as it is when $\Delta=t$ and $\mu=0$, the absorption spectrum has an even and an odd part pertaining to each ground state
\begin{align}
        A_{(i)}(E) &= |\beta_{\rm even}|^2\sum_{\phi_{\rm odd}} |\langle\phi_{\rm odd} |P_{(i)}\ket{\rm GS_{\rm even}}|^2 \delta(E-E_{\phi_{\rm odd}} + E_{\rm GS}) \nonumber \\
        &+ |\beta_{\rm odd}|^2\sum_{\phi_{\rm even}} |\langle\phi_{\rm even} |P_{(i)}\ket{\rm GS_{\rm odd}}|^2 \delta(E-E_{\phi_{\rm even}} + E_{\rm GS})\nonumber\\
        &= |\beta_{\rm even}|^2 A_{(i)}^{\rm even}(E) + |\beta_{\rm odd}|^2 A_{(i)}^{\rm odd}(E),
        \label{eq:degenerate_absorption}
\end{align}
where $\ket{\phi_{\rm even(odd)}}$ are the eigenstates of the one hole subspace and the corresponding electron parity, and we used the notation $A_{(i)}$ to simultaneously refer to the regular absorption spectrum $A$, and $A_i$ the spatially resolved absorption spectrum coming from dot $i$.

\subsubsection{Analytic result for localized hole \label{sec:absorption_local_hole}}
If $\tau=0$ and the hole is localized the full Hamiltonian becomes block diagonal (see \eqref{eq:H_matrix_hole_structure_N3}), i.e. the subspaces of having the hole in each of the dots decouple. Consequently, we have
\begin{equation}
 A(E) = \frac1N \sum_{i=1}^N A_i(E)   .
 \label{eq:A_sum_Ai}
\end{equation}

\begin{figure}[H]
    \centering
	\includegraphics[width=0.8\columnwidth]{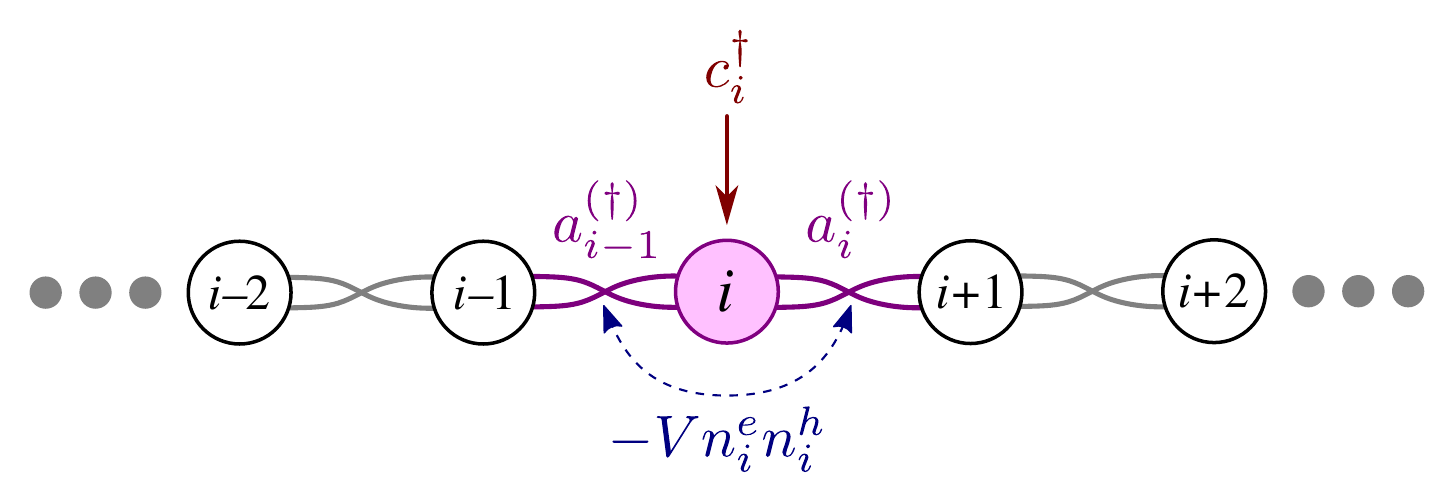}
	\caption{An electron created by $c_i^\dagger$ is a superposition of creation and annihilation operators of two bond Fermions $a_i^{(\dagger)}$ and $a_{i-1}^{(\dagger)}$, according to \eqref{eq:c_to_a}. The interaction $-Vn_i^en_i^h$ mixes up the two bond Fermions according to \eqref{eq:ne_j_mixing}. Note that when $i$ is one of the two ends, then one of the bond Fermions is the zero mode $a_N$ (see the Appendix~\ref{sec:appendix} for more details).}
	\label{fig:abs_mechanism}
\end{figure}

At the heart of topological regime when $\Delta=t$ and $\mu=0$, as depicted graphically in Fig.~\ref{fig:abs_mechanism}, and expressed in \eqref{eq:c_to_a} , 
an electron created at site $i$ by $c_i^\dagger$, decomposes into a superposition of creation and annihilation operators of the two bond Fermions on its two sides, $a_i^{(\dagger)}$ and $a_{i-1}^{(\dagger)}$. If the electron is created at one of the two ends, one of the bond Fermions is the zero mode $a_N$.
On the other hand, when the hole is at site $i$, the interaction $-Vn_i^en_i^h$ mixes the two bond Fermions, as shown in Fig.~\ref{fig:abs_mechanism}, and expressed in Eqs.~(\ref{eq:ne_j_mixing} and \ref{eq:ne_1_mixing}).
 As we show in the Appendix~\ref{sec:appendix}, combining these two mechanisms, one can find an analytic expression for the spatially resolved absorption spectrum for a chain of arbitrary length $N$ if the hole is created on site $i$ as
    \begin{equation}
	A_i(E) = \frac{1}{2}
	\begin{cases}
		\delta(E - E_0) & 1<i<N \\
        A_-\delta(E-E_-) +
        A_+\delta(E-E_+) & i=1,N
	\end{cases},
    \label{eq:Ai_special_final_result}
    \end{equation}
where
\begin{subequations}
    \begin{align}
        \label{eq:E0}
        & E_0 = \eta + 2|t| - V,\\
        \label{eq:Epm}
        & E_\pm = \eta + |t| - \frac V2 \pm \sqrt{t^2 + \left(\frac V2\right)^2},\\
        \label{eq:A_pm}
       & A_\pm = \frac12 \left( 1 \mp \frac{V}{\sqrt{4t^2 + V^2}}\right),
    \end{align}
    \label{eq:E_A_final_result}
\end{subequations}
and then the full absorption spectrum is given by the simple sum in \eqref{eq:A_sum_Ai}.
\begin{figure}[h]
\centering
\includegraphics[width=\columnwidth]{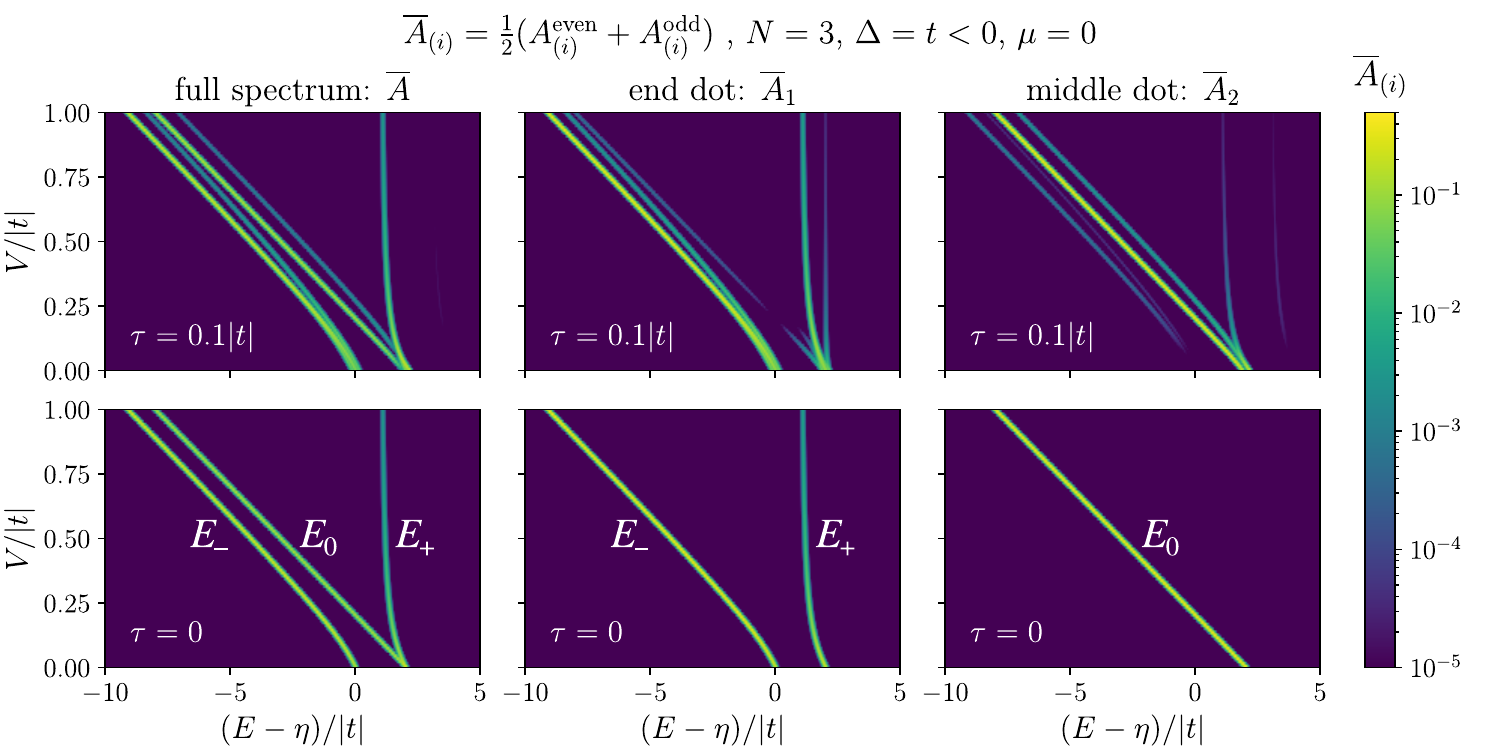}
\caption{(left) The averaged absorption spectrum $\overline{A}(E)$, and (middle and right) spatially resolved absorption $\overline{A}_i(E)$, for $\Delta=t,\mu=0$, and for $N=3$ dots. (top) for a mobile hole with $\tau=0.1|t|$, (bottom) for a localized hole, $\tau=0$, according to the analytic results in Eqs.~(\ref{eq:Ai_special_final_result} and \ref{eq:E_A_final_result}).
The spectra are plotted against $(E-\eta)/|t|$ while changing $V/|t|$ on the y-axis.
    The bright curves show the location of the peaks as $V$ changes, and the color scale shows their heights. 
    Gaussian profile was used for the peaks with the width $\sigma = 0.025|t|$. The maximum value of each peak shows the magnitude of the corresponding matrix element.}
\label{fig:absorption_bond_EV_tau_0.1}
\end{figure}
The bottom row of Figure~\ref{fig:absorption_bond_EV_tau_0.1} shows the results in Eqs.~(\ref{eq:Ai_special_final_result} and \ref{eq:E_A_final_result}) for the case of $N=3$.
The peak $E_0$ is only present in the middle, while the peaks $E_\pm$ are present at the two ends of the chain.
As we show in the Appendix~\ref{sec:appendix}, the two peaks $E_\pm$ have a mixture of zero mode in them, while $E_0$ is purely made of non-zero bond Fermions.
At $V=0$, $E_-$ is purely made of zero mode while $E_+$ is purely made of non-zero bond Fermions. As we increase $V$, $E_+$ acquires more zero mode contribution while $E_-$ mixes more with a non-zero bond Fermions.
At the same time, by increasing $V$, the peak at $E_+$ diminishes, as can be seen from \eqref{eq:A_pm}.
If not too weak, $E_+$ peak is a better resolved optical signature for the MZM than $E_-$, as it is separated from the rest of the spectrum by $V$, and we expect to have $V\gg |t|$. This presents an advantage over scanning tunnelling microscopy approach for detecting MZM \cite{jack2021detecting}. 
Moreover, if one can perform spatially resolved absorption spectroscopy on the chain, the presence of the zero mode can be determined by the presence of a visible peak at high energy near $E_+$ when probing the end dots, and its absence when probing other dots.

\subsubsection{Absorption for mobile hole \label{sec:absorption_mobile_hole}}

When the hole is mobile, there are $N$ itinerant hole states with different energies. Therefore, one would expect $N$ different transitions to each electronic state.
More importantly, as can be seen in \eqref{eq:H_matrix_hole_structure_N3}, hopping hole mixes up different subspaces of having the hole in different dots. As a result more transitions become allowed leading to the emergence of more peaks in the absorption spectrum.

In Fig.~\ref{fig:absorption_bond_EV_tau_0.1} we  compare the absorption spectrum of a mobile hole with $\tau=0.1|t|$ (top row), and the analytic result of \eqref{eq:Ai_special_final_result} for localized hole (bottom row), for the case of $N=3$.
It is evident how more peaks are visible for the case of mobile hole, while the major peaks are still close to the location of $E_0$ and $E_\pm$.
Moreover, note how in the full spectrum $\overline{A}$ (top left) there is only one visible peak at high energy near $E_+$, and how the same peak is large in $\overline{A}_1$ (top middle) and faint in $\overline{A}_2$ (top right), pertaining to the localized nature of the MZM that $E_+$ carries.
In plotting Fig.~\ref{fig:absorption_bond_EV_tau_0.1} we used $\overline{A}_{(i)} = \frac12 (A^{\rm even}_{(i)} + A^{\rm odd}_{(i)})$, as for a mobile hole the even and odd parts of the absorption spectrum are not the same. But since there is no preference between the two ground states, one would expect to observe an average of the two.

\begin{figure}[h]
\centering
\includegraphics[width=\columnwidth]{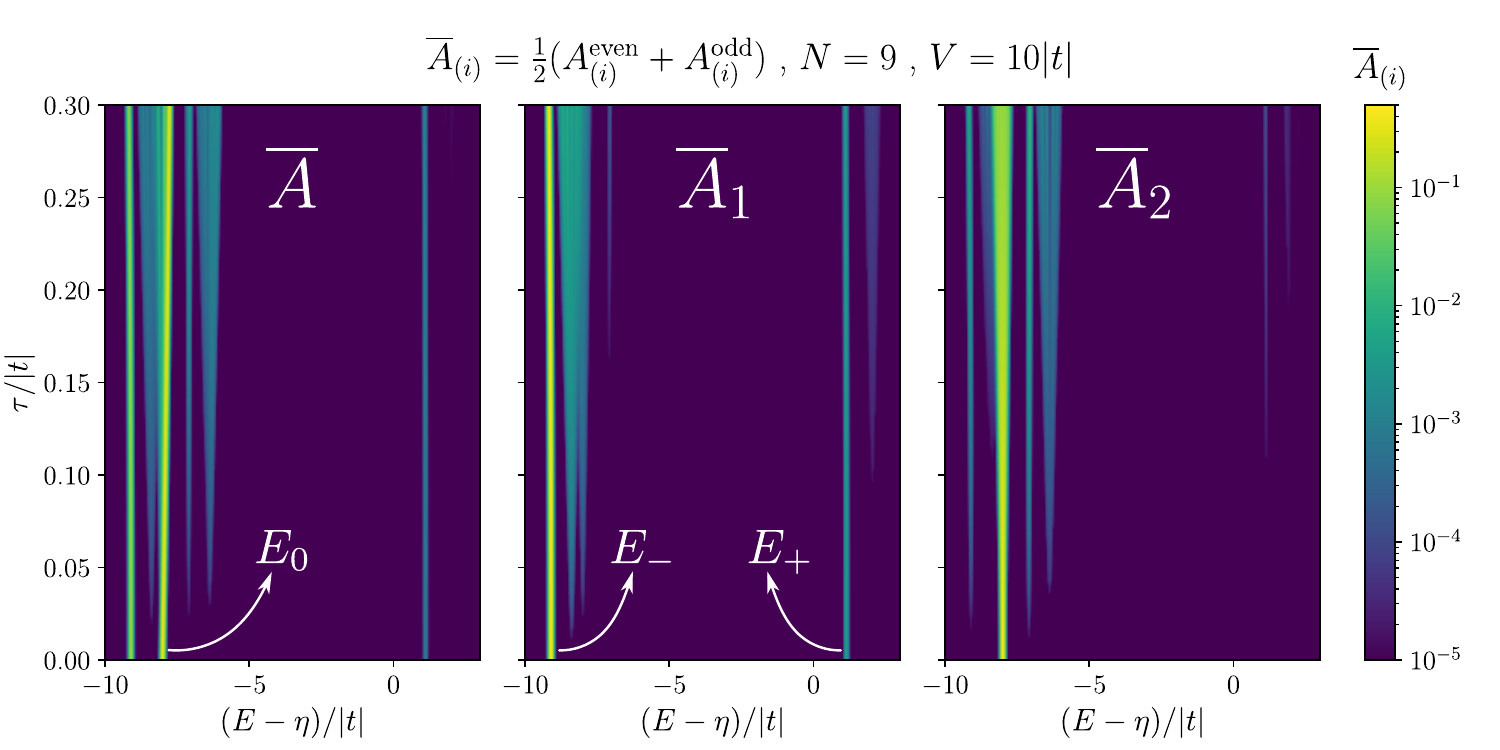}
\caption{Absorption spectrum for a chain of length $N=9$, and for $\Delta=t,\mu=0, V=10|t|$, and changing $\tau$. (left) The full averaged spectrum $\overline{A}$, (middle) the spatially resolved spectrum for the first dot $\overline{A}_1$, and (right) the spatially resolved spectrum for the second dot $\overline{A}_2$.
    The bright curves show the location of the peaks as $\tau$ changes, and the colorscale shows their heights.     
    Gaussian profile was used for the peaks with the width $\sigma = 0.025|t|$. The maximum value of each peak shows the magnitude of the corresponding matrix element.}
\label{fig:absorption_long_chain}
\end{figure}
The same logic is valid for a chain of any length, as the analytic result in \eqref{eq:Ai_special_final_result} is for general $N$.
Figure~\ref{fig:absorption_long_chain} shows the absorption spectrum of a chain of length $N=9$.
Here, we set $V=10|t|$ while changing $\tau$.
When $\tau\to0$ we approach the idealized case of localized hole, where the subspaces of having the hole on each dot are decoupled. Growing $\tau$ mixes up the modes of different dots.
Consequently, the zero mode starts leaking out of the two ends of the chain.
On the first panel of Fig.~\ref{fig:absorption_long_chain} we can see that at high energy there is still only one visible peak near $E_+$, until about $\tau=0.3|t|$ a faint peak appears to the right of it.
This makes $E_+$ a very robust signature for a relatively large range of hole hopping.
Having access to spatially resolved spectrum we can further confirm that the peak is indeed coming from the two ends.
It can be seen from the third panel of Fig.~\ref{fig:absorption_long_chain} that the there is no visible high energy peak on the site next to the end dot ($\overline{A}_2$) until around $\tau=0.1|t|$.
In contrast, we can observe in $\overline{A}_2$ that $E_-$, which also contains a large share of zero mode and is a stronger peak, starts leaking out of the end dot very quickly for small $\tau$'s.

\section{Conclusion \label{sec:conclusion}}
We present here a theory of Majorana excitons, photo-excited  conduction electron-valence band hole pairs, interacting with Majorana Fermions in a Kitaev chain of semiconductor quantum dots embedded in a  nanowire. Using exact diagonalization techniques and Majorana and Bond Fermions we  
 compute the energy spectra of the system. We confirm the existence of nonlocal bond Fermion, a superposition of Majorana Fermions at the two ends of the chain,  with zero energy. We  introduce a valence band hole and describe its interaction with Majorana fermions. We predict interband absorption spectra and discuss the signature of Majorana Zero Modes in the absorption spectra. We demonstrate how spatially resolved absorption spectrum can be used to confirm the localized character of the MZMs.

We hope this preliminary work  motivates future theoretical and experimental work on hybrid nanowire semiconductor quantum dots /superconductor systems for the demonstration of Majorana Fermions. 

\begin{acknowledgments}
This research  was supported by the Quantum Sensors and Applied Quantum Computing Challenge Programs at the National Research Council of Canada, by NSERC Discovery Grant No. RGPIN- 2019-05714, and University of Ottawa Research Chair in Quantum Theory of Quantum Materials, Nanostructures, and Devices.
\end{acknowledgments}

\appendix
\section{Exact diagonalization for chain of length three \label{sec:N3}}

Following \eqref{eq:kitaev_confs}, for a chain of length $N=3$, the configurations of normal Fermions in the even and odd subspaces are
\begin{subequations}
    \begin{align}
        \label{eq:3qd_confs_even}
        \text{Even:} \  &\lbrace \ket{000} \ , \  \ket{110} \ , \ \ket{101} \ , \ \ket{011}\rbrace,\\
        \label{eq:3qd_confs_odd}
        \text{Odd:} \ &\lbrace \ket{100} \ , \  \ket{010} \ , \ \ket{001} \ , \ \ket{111}\rbrace.
    \end{align}
    \label{eq:3qd_confs}
\end{subequations}
Computing the matrix elements of all terms in the Kitaev Hamiltonian $H_e$ in \eqref{eq:Kitaev_Ham} between every pairs of configurations in \eqref{eq:3qd_confs_even} and \eqref{eq:3qd_confs_odd} we can explicitly write the matrix $\bra{q_{M'},M'} H_e \ket{p_M,M}$ in \eqref{eq:Wfn_MatrixH}, in each of the subspaces as
\begin{subequations}
    \begin{align}
      \label{eq:3Site_MatrixH_even}  
      H_{c}^{\text{even}} &= \begin{bmatrix} 0 & -\Delta & 0 & -\Delta
      \\ -\Delta & -2\mu & t & 0
      \\0 & t & -2\mu & t 
      \\ -\Delta & 0 & t & -2\mu \end{bmatrix},\\[15pt]
     H_{c}^{\text{odd}} &= \begin{bmatrix} -\mu & t & 0 & -\Delta
      \\ t & -\mu & t & 0
      \\0 & t & -\mu & -\Delta 
      \\ -\Delta & 0 & -\Delta & -3\mu \end{bmatrix},
     \label{eq:3Site_MatrixH_odd}
    \end{align}
    \label{eq:3qd_H_matrices}
\end{subequations}
for which we used the same ordering as in \eqref{eq:3qd_confs}.

Then, for ED in the bond Fermion basis, following \eqref{eq:bond_confs}, the configurations are
\begin{subequations}
    \begin{align}
        \label{eq:3qd_confs_even_bond}
        \text{Even:} \  &\lbrace \ket{\overline{000}} \ , \  \ket{\overline{110}} \ , \ \ket{\overline{101}} \ , \ \ket{\overline{011}}\rbrace,\\
        \label{eq:3qd_confs_odd_bond}
        \text{Odd:} \ &\lbrace \ket{\overline{100}} \ , \  \ket{\overline{010}} \ , \ \ket{\overline{001}} \ , \ \ket{\overline{111}}\rbrace,
    \end{align}
    \label{eq:3qd_confs_bond}
\end{subequations}
and one can use \eqref{eq:General_H_BondFermions} to find the corresponding Kitaev Hamiltonian matrices in each of the subspaces as
\begin{subequations}
\begin{align}
   & H_{a}^{\text{even}} = \frac12
   \begin{bmatrix}
        -2(t +\Delta) & -\mu & t -\Delta + \mu & \mu - t +\Delta \\[5pt]
        -\mu & 2(t +\Delta) & t -\Delta - \mu & \mu - t +\Delta \\[5pt]
        t -\Delta + \mu & t -\Delta - \mu & 0 & -\mu \\[5pt]
        \mu - t +\Delta & \mu - t +\Delta & -\mu & 0
      \end{bmatrix} - \frac32\mu,
     \label{eq:MatrixH_BF_tDeltamueven} \\[15pt]
     & H_{a}^{\text{odd}} = \frac12
     \begin{bmatrix}
        0 & -\mu & t -\Delta - \mu & \Delta-t-\mu \\[5pt]
        -\mu & 0 & t -\Delta - \mu & \Delta-t-\mu \\[5pt]
        t -\Delta - \mu & t -\Delta - \mu & -2(t+\Delta) & -\mu \\[5pt]
        \Delta-t-\mu & \Delta-t-\mu & -\mu & 2(t+\Delta)
      \end{bmatrix} - \frac32\mu,
     \label{eq:MatrixH_BF_tDeltamuodd}
\end{align}
\label{eq:H_Matrix_bond}
\end{subequations}
for which we used the same ordering as in \eqref{eq:3qd_confs_bond}.
Notice how the two Hamiltonian matrices in \eqref{eq:H_Matrix_bond} become diagonal when $\Delta=t$ and $\mu=0$.

\section{Analytic calculation of absorption spectrum for localized hole \label{sec:appendix}}
Following the notation in Table~\ref{tab:elec_energy_spectrum}, for a chain of arbitrary length $N$, and for $\Delta=t<0$ and $\mu=0$, we express the two degenerate ground states of the system as
\begin{subequations}
    \begin{align}
        \label{eq:GS1}
        &\ket{\rm GS} = \prod_{j=1}^{N} a_j^\dagger |0_a\rangle, \\
        \label{eq:GS2}
        &\ket{\overline{\rm GS}} = \prod_{j=1}^{N-1} a_j^\dagger |0_a\rangle.
    \end{align}
    \label{eq:GS_special}
\end{subequations}
Here, we derive the result in Eqs.~(\ref{eq:Ai_special_final_result} and \ref{eq:E_A_final_result}) using $\ket{\rm GS}$ in \eqref{eq:GS1}, and the procedure is the same for $\ket{\overline{\rm GS}}$.

To start, first note that from \eqref{eq:a_to_c} we have
\begin{subequations}
    \begin{align}
        \label{eq:c_j_to_a_j}
        c_j &= \frac{1}{2}(a_j^\dagger + a_j + a_{j-1} - a_{j-1}^\dagger),
        & 1 < j \leq N \\
        \label{eq:c1_to_a}
        c_1 &= \frac{1}{2}(a_1^\dagger + a_1 + a_{N} - a_{N}^\dagger),
        & j=1.
    \end{align}
    \label{eq:c_to_a}
\end{subequations}
which means for $1<j<N$ we have
\begin{equation}
    P_j\ket{\rm GS} =
    c_j^\dagger h_j^\dagger\ket{\rm GS} =
    \frac{1}{2}(a_j - a_{j-1})\ket{{\rm GS};j}=
    \frac{1}{2}(\ket{a_j;j} - \ket{a_{j-1};j}),
    \label{eq:Pj_GS}
\end{equation}
where we used the same notation as in Table~\ref{tab:bond_hole_configs} for the excited states.
On the other hand, from \eqref{eq:ne_j_to_a_j} we have
\begin{equation}
     n_j^e
        \begin{pmatrix}
            \ket{a_{j-1}} \\
            \ket{a_{j}} \\
        \end{pmatrix}=\frac12
        \begin{pmatrix}
            1 & -1 \\
            -1 & 1 \\
        \end{pmatrix}
        \begin{pmatrix}
            \ket{a_{j-1}} \\
            \ket{a_{j}} \\
        \end{pmatrix}.
        \label{eq:ne_j_mixing}        
\end{equation}
Recalling that both $\ket{a_j}$ and $\ket{a_{j-1}}$ are eigenstates of $H_e$ with excitation energy $2|t|$, then in the basis $\lbrace\ket{a_j; j}, \ket{a_{j-1}; j}\rbrace$, the full Hamiltonian $H=H_e+\eta-Vn_j^e$ is
    \begin{equation}
    H = 2|t|+\eta -\frac V2 
        \begin{pmatrix}
            1 & -1 \\
            -1 & 1 \\
        \end{pmatrix},
        \label{eq:full_H_aj_mid}
    \end{equation}
with the following two eigenstates
    \begin{subequations}
        \begin{align}
            \label{eq:GSj-}
            \ket{a_{j}^-; j} &= \frac{1}{\sqrt{2}} \left(
            \ket{a_{j}; j} - \ket{a_{j-1}; j}
            \right),
            & E_0 = \eta + 2|t| - V \\
            \label{eq:GSj+}
            \ket{a_{j}^+; j} &= \frac{1}{\sqrt{2}} \left(
            \ket{a_{j}; j} + \ket{a_{j-1}; j}
            \right),
            & E_1 = \eta + 2|t|.
        \end{align}
        \label{eq:GSj_pm}
    \end{subequations}
Now using \eqref{eq:Pj_GS} we have $\left|\langle a_{j}^-; j| P_j \ket{\rm GS}\right|^2 = \frac12$, and $\left|\langle a_{j}^+; j | P_j \ket{\rm GS}\right|^2 = 0$, which give us the result in \eqref{eq:Ai_special_final_result} for the case of $1<j<N$.

Next, we show the second case for $j=1$, and the procedure is the same for $j=N$.
In this case, using \eqref{eq:c1_to_a} we have
\begin{equation}
    P_1\ket{\rm GS} =
    c_1^\dagger h_1^\dagger\ket{\rm GS} =
    \frac{1}{2}(a_1 - a_{N})\ket{{\rm GS};1}=
    \frac{1}{2}(\ket{a_1;1} - \ket{a_N;1}),
    \label{eq:P1_GS}
\end{equation}
where $\ket{a_N}$ is identical to the other ground state $\ket{\overline{\rm GS}}$ up to a global phase, hence its excitation energy is zero. Then from \eqref{eq:ne1_to_a} we have
\begin{equation}
    n_1^e
        \begin{pmatrix}
            \ket{a_N} \\
            \ket{a_1} \\
        \end{pmatrix}=\frac12
        \begin{pmatrix}
            1 & -1 \\
            -1 & 1 \\
        \end{pmatrix}
        \begin{pmatrix}
            \ket{a_N} \\
            \ket{a_1} \\
        \end{pmatrix}.
        \label{eq:ne_1_mixing}
\end{equation}
Therefore, considering that $\ket{a_1}$ and $\ket{a_N}$ are eigenstates of $H_e$ with excitation energy $2|t|$ and zero, respectively,  the full Hamiltonian $H=H_e+\eta-Vn_1^e$ in the basis $\lbrace\ket{a_N; 1}, \ket{a_1; 1}\rbrace$ is
\begin{equation}
H = \eta -\frac V2 + 
    \begin{pmatrix}
        0 & \frac V2 \\
        \frac V2 & 2|t| \\
    \end{pmatrix},
    \label{eq:full_H_a1}
\end{equation}
and its two eigenstates are given by
\begin{subequations}
    \begin{align}
        \label{eq:a1-}
        &\ket{a_1^-; 1} = 
        \cos(\theta)\ket{a_1; 1} - \sin(\theta)\ket{a_N; 1}            ,
        & E_- = \eta + |t| - \frac V2  - \sqrt{t^2 + \left(\frac V2\right)^2},\\
        \label{eq:a1+}
        &\ket{a_1^+; 1} = 
        \sin(\theta)\ket{a_1; 1} + \cos(\theta)\ket{a_N; 1},
        & E_+ = \eta + |t| - \frac V2 + \sqrt{t^2 + \left(\frac V2\right)^2},
    \end{align}
    \label{eq:a1_pm}
\end{subequations}
with
\begin{align}
    &\cos(\theta) = \sqrt{\frac 12 + \frac{t^2}{\sqrt{4t^2+V^2}}},
    &\sin(\theta) = \sqrt{\frac 12 - \frac{t^2}{\sqrt{4t^2+V^2}}}.
    \label{eq:theta}
\end{align}
Now using \eqref{eq:P1_GS} we have $\left|\bra{a_{1}^-; 1} P_1 \ket{\rm GS}\right|^2 = \frac12(\cos(\theta)+\sin(\theta))^2=\frac12 A_-$, and $\left|\bra{a_{1}^+; 1} P_1 \ket{\rm GS}\right|^2 = \frac12(\cos(\theta)-\sin(\theta))^2=\frac12 A_+$, where we used \eqref{eq:theta} to express these matrix elements in terms of $A_\pm$ in \eqref{eq:A_pm}. From this follows the result in \eqref{eq:Ai_special_final_result} for the case of $j=1,N$.

\bibliographystyle{apsrev4-2}
\bibliography{ref}

\end{document}